\documentclass[11pt,a4paper]{article}
\usepackage[dvips]{graphicx}
\usepackage{amsfonts}
\begin{document}
\thispagestyle{empty}
\renewcommand{\baselinestretch}{1.2} 
\small\normalsize
\frenchspacing
\noindent
{\Large \textbf{How  quantum mechanics with deterministic collapse\\[3pt]localizes 
 macroscopic objects}}
\\
\\
{\bf{Arthur Jabs}}
\renewcommand{\baselinestretch}{1}
\small\normalsize
\\
\\
Alumnus, Technical University Berlin. 

\noindent
Vo\ss str. 9, 10117 Berlin, Germany 

\noindent
arthur.jabs@alumni.tu-berlin.de
\\
\\
(31 July 2019)

\newcommand{\rmi}{\mathrm{i}}
\vspace{15pt}

\noindent
{\bf{Abstract.}} Why microscopic objects exhibit wave properties (are delocalized), but macroscopic do not (are localized)? Traditional quantum mechanics attributes wave properties to all objects. When complemented with a deterministic collapse model (Quantum Stud.: Math. Found. {\bf3}, 379 (2016)) quantum mechanics can dissolve the discrepancy. Collapse in this model means contraction and occurs when the object gets in touch with other objects and satisfies a certain criterion. One single collapse usually does not suffice for localization. But  the object rapidly gets in touch with other objects in a short time, leading to rapid localization. Decoherence is  not involved. 

\begin{list}
{\textbf{Keywords:}}
{\setlength{\labelwidth}{2.0cm} 
 \setlength{\leftmargin}{2.2cm}
 \setlength{\labelsep}{0.2cm} }
\item
microscopic/macroscopic transition, superposition, deterministic collapse model
\end{list}

\newcommand{\rmf}{\mathrm{f}}
\newcommand{\rmd}{\mathrm{d}} 
\newcommand{\sbl}{\hspace{1pt}}
\newcommand{\bfitp}{\emph{\boldmath $p$}}
\newcommand{\bfitr}{\emph{\boldmath $r$}}
\newcommand{\bfitsr}{\emph{\footnotesize{\boldmath $r$}}}
\newcommand{\bfitQ}{\emph{\boldmath $Q$}}
\newcommand{\bfitk}{\emph{\boldmath $k$}}
\newcommand{\PSI}[1]{\Psi_{\textrm{\footnotesize{#1}}}}
\newcommand{\PHI}[1]{\Phi_{\textrm{\footnotesize{#1}}}}
\newcommand{\spsi}[1]{\psi_{\textrm{\footnotesize{#1}}}}
\newcommand{\pcop}{P_{\textrm{\footnotesize{Cop}}}}
\newcommand{\pred}{P_{\textrm{\footnotesize{red}}}}
\newcommand{\psis}{\psi_{\rm s}(\bfitr,t)}

\vspace{20pt}
\noindent
{\bf{1  The deterministic collapse model}}
\bigskip

\noindent
The conclusions of the present note are consequences of the deterministic collapse model [1]. We therefore briefly recall those features that are required here. Thus, collapse occurs when two wavepackets, representing microscopic or macroscopic objects, overlap and satisfy the following criterion:

\begin{displaymath} 
\hspace{150pt}
 |\alpha_1-\alpha_2| \leq \mbox{{\normalsize{$\frac{1}{2}$}}}\,\alpha_{\textrm s} \hspace{153pt}(1)
\end{displaymath} 

\vspace{-10pt}

\begin{displaymath}
\hspace{108pt} 
\bigg[\int_{\mathbb{R}^3} |\spsi{1}(\bfitr,t)| \, |\spsi{2}(\bfitr,t)| \, {\textrm d}^3r\bigg]^2 \ge \alpha/{2\pi}.
\hspace{92pt}(2)
\end{displaymath}

\vspace{8pt}

\noindent
$\alpha_ 1$ is the absolute phase constant of wavepacket $\psi_1$, and $\alpha_ 2$ that of $\psi_2$. These constants are new elements of the model, and are  pseudorandom numbers in the interval $[0,2\pi]$ modulo $2\pi$. $\alpha_{\rm S}$ is Sommerfeld's fine-structure constant. $ \alpha$ is the smaller of $\alpha_1$ and $\alpha_2$

The collapse, then, suddenly contracts both wavepackets to the overlap volume, that is, where $|\psi_1(\bfitr,t)|\,  |\psi_2(\bfitr,t)|$ is practically concentrated (its effective support). According to the formulas (1), (2) the overlap volume need not be extremely small.

\vspace{70pt}
\noindent
{\bf{2  Quantum mechanical objects}}
\smallskip

\noindent
We consider nonrelativistic quantum mechanics and  describe an object by the wavepacket:
\begin{displaymath}
\hspace{105pt} \Psi=e^{\rmi \alpha}\,\psi(\bfitr,t)\times \psi_{\rm R}(\rho_1,\cdots,\rho_N,t).\hspace{120pt}(3)
\end{displaymath}
\noindent
$e^{\rmi\alpha}\psi(\bfitr,t)$ is the
center-of-mass (CM) function, which is a superposition of de Broglie waves representing the free object as a whole. $\psi_{\rm R}(\rho_1,\cdots,\rho_N,t)$  is the internal function, which represents  the relative positions $\rho_i$ and the internal dynamics of the elementary particles or clusters constituting the object [2].  For an elementary particle, there is only a CM function. The width (effective support, spatial volume) of $ |\psi_{\rm R}(\rho_1,\cdots,\rho_N,t)|^2$ represents the size of the object. The spatial volume of the CM function may be much larger than that of the internal function. When the volume of the CM function is very small, say that of an atom, the objet is called localized,  otherwise delocalized.

A microscopic object of mass $1.7 \times 10^{-23}$ kg (molecule of tetraphenylporphyrin) and diameter $5\times 10^{-9}$ m ($\lambda_{\rm{deBr}}=4\times 10^{-12} $ m with $v_0=10$ m/s) can be delocalized  over a hundred times its own diameter [3]. A macroscopic object which can be seen, touched, and tasted like a grain of sugar of mass $10^{-7}$ kg and diameter $0.5\times 10^{-3}$ m  ($\lambda_{\rm{deBr}}=7\times 10^{-28} $ m with $v_0=10$ m/s)  is always observed to be localized. Why?

\vspace{20pt}
\noindent
{\bf{3  Localization}}
\smallskip

\noindent
Consider a particular delocalized object. When its CM function overlaps with the function of another object and the criterion for collapse (1), (2) is satisfied, the volume of the CM function of our object (as well as that of the other object) contracts to the overlap volume. This volume may be relatively large, so that this collapse does not succeed in localizing our object. However, any subsequent collapse cannot enlarge the volume of the object's CM function, only diminish it. Now, an object suffers many collapses in a short time due to the many other objects (photons, air molecules, etc.) in its environment, and these rapidly localize the object.
 \vspace{5pt}
 
It is reasonable to assume that the considered object's phase constant, say $\alpha_1$,  which enters formula (1), is that of its CM function $\psi$, as long as the volume of  $\psi$ totally covers the volume of the internal function $\psi_{\rm R}$. If this ceases to be the case in the process of localization, some objects from the environment may no longer overlap with the CM function $\psi$, but only with the wave function of one of the clusters, which constitute the object [4]. That is, $\alpha_1$ is no longer the phase constant of the CM function $\psi$, but that of one of the clusters. This decreases the shrinking rate of  the volume of $\psi$, that is, of its final localization. Due to the large number of environmental objects, however, the rate will still be extremely high.
 
\vspace{70pt}
\noindent
{\bf{4  Transition micro-macro}}
\bigskip

\noindent
 So far the above considerations apply to both macroscopic and microscopic objects. Imagine that both move in the same environment. Now the question is reversed: why do  microscopic objects remain delocalized? The answer lies in the spreading of a wavepacket due to Schr\"odinger dynamics. This spreading velocity $v_{\rm S}$ (transverse as well as longitudinal) is given by [5]:
\[ 
\hspace{155pt}
v_{\rm S}=\frac{\hbar}{d\; m_0}. \hspace{170pt}(4)
\]
$d$ is the minimum diameter of the object at the beginning of spreading, and $m_0$ is its mass.

Let us consider the molecule of tetraphenylporphyrin mentioned in Sec. 2 as a microscopic object. Let us assume that its minimum radius is of the order of the Bohr radius $5\times 10^{-11}$ m. Then its spreading velocity is $v_{\rm Smic}=$ 6 cm/s .

Let us, on the other hand, take the grain of sugar mentioned in Sec. 2 as a macroscopic object, and let its minimum radius again be  $5\times 10^{-11}$ m. Then $v_{\rm Smac}=10^{-17}\, {\rm m/s}=3\times 10^{-10}\, {\rm m/year}$.

These examples demonstrate that after a contraction due to collapse microscopic objects rapidly recover their delocalization, whereas macroscopic objects cannot because their spreading velocity is negligible.

\vspace{5pt}
So, somewhere between tetraphenylporphyrin molecules and grains of sugar lies the borderline between microscopic and macroscopic objects. Actually, it is difficult, if not impossible, to exactly define it because it depends on the environment [3, p. 9]. This is in line with the observation that  even the delocalization of microscopic objects lasts only for limited time intervals [3, p. 2, 3, 6, 9].  In any case, mass plays an important role  because it determines the spreading velocity.

\vspace{5pt}
This provides also the justification of the usual assertion that macroscopic objects show no wave properties because their de Broglie wavelengths are so small:  
both   $v_{\rm S}$ of Eq. (4) and $\lambda$ in the de Broglie relation $\lambda=h/m_0v_0$   ($v_0$ = velocity of the object's center) are proportional to $\hbar$. Thus, simple algebra gives us the proportionality between $v_{\rm S}$ and $\lambda$ in the form:
\vspace{-5pt}
\[ 
\hspace{170pt}
v_{\rm S}=\lambda \frac{v_0}{2\pi d}. \hspace{155pt}(5)
\]
\vspace{10pt} 

\noindent
{\textbf{Notes and References}} 
\begin{enumerate}
\renewcommand{\labelenumi}{[\arabic{enumi}]}
\hyphenpenalty=1000

\item Jabs, A.: A conjecture concerning determinism, reduction, and measurement in quantum mechanics, arXiv:1204.0614 (2019) (Quantum Stud.: Math. Found. {\bf3} (4), 279-292 (2016))

\item Messiah, A.: Quantum Mechanics (North-Holland Publishing Company, \mbox{Amsterdam}, 1970)  Chapter IX, \S \, 12, 13

\item Arndt, M. and Hornberger, K.: Testing the limits of quantum mechanical superpositions, arXiv:1410.0270 (Nature Physics {\bf10}, 271-277 (2014) p. 4, 6)

\item This is the effect that resolves the `measurement problem', as expounded in [1]

\item Jabs, A.: Quantum mechanics in terms of realism, arXiv:quant-ph/9606017 (2019) Appendix  A

\end{enumerate}
\bigskip
\hspace{5cm}
------------------------------
\end{document}